# Lightweight Convolutional Neural Networks for Retinal Disease Classification

1st Duaa Kareem Qasim
dept of Computer Techniques Engineering
Imam Alkadhim University College
Baghdad, Iraq
2101143069@alkadhum-col.edu.iq

2nd Sabah Abdulazeez Jebur
dept of Cyber Security
Imam Alkadhim University College
Baghdad, Iraq
sabah.abdulazeez@iku.edu.iq

3th Lafta Raheem Ali
General Directorate of Education of Salahuddin
Salahuddin, Iraq
l.alkhazraji@gmail.com

4th Abdul Jalil M. Khalaf
Department of Cybersecurity
Imam AlKadhim University College
Baghdad, Iraq
amkhalaf@iku.edu.iq

5th Abir Jaafar Hussain
Department of Electrical Engineering,
University of Sharjah, Sharjah, United Arab Emirates
abir.hussain@sharjah.ac.ae

*Abstract*— Retinal diseases such as Diabetic Retinopathy (DR) and Macular Hole (MH) significantly impact vision and affect millions worldwide. Early detection is crucial, as DR, a complication of diabetes, damages retinal blood vessels, potentially leading to blindness, while MH disrupts central vision, affecting tasks like reading and facial recognition. This paper employed two lightweight and efficient Convolution Neural Network architectures, MobileNet and NASNetMobile, for the classification of Normal, DR, and MH retinal images. The models were trained on the RFMiD dataset, consisting of 3,200 fundus images, after undergoing preprocessing steps such as resizing, normalization, and augmentation. To address data scarcity, this study leveraged transfer learning and data augmentation techniques, enhancing model generalization and performance. The experimental results demonstrate that MobileNetV2 achieved the highest accuracy of 90.8%, outperforming NASNetMobile, which achieved 89.5% accuracy. These findings highlight the effectiveness of CNNs in retinal disease classification, providing a foundation for AI-assisted ophthalmic diagnosis and early intervention.

## I. INTRODUCTION

Retinal diseases of many kinds affect millions of people all over the world, but perhaps two of the most common and vital ones are Diabetic Retinopathy (DR) and Macular Hole [1]. Diabetic Retinopathy is a complication of diabetes that damages blood vessels in the retina and this could lead to vision loss if left untreated, the early stages of DR are often asymptomatic, making regular screening essential for individuals with diabetes.

On the other side. Macular Hole is a defect in the central part of the retina which is known as the macula, which is responsible for clear central vision. A macular hole can cause severe visual impairment edit particularly affects activities like reading texts or recognizing people [2].

Convolutional Neural Networks (CNNs) have greatly impacted the field of computer vision, particularly in the field of medical image analysis where they have demonstrated exceptional performance in detecting and diagnosing various diseases [3]. CNNs have the potential to assist ophthalmologists in making quicker and more accurate diagnoses, and ultimately leads to better outcomes for patients [4].

CNNs identify complex patterns by progressively combining simpler ones through specialized layers. They consist of an input layer, multiple hidden layers (convolutional, pooling, or fully connected), and an output layer [4], [5]. The convolutional layer, which is the core part of a process, implements learnable filters with small receptive fields that cover the entire input depth. In the forward pass, these filters shift across the input calculating dot products to create activation maps which allow sophisticated features in retinal images to be detected [6].

Training CNNs for optimal performance typically requires large datasets, which presents a significant challenge [7]. To overcome this, transfer learning (TL) has gained widespread acceptance among researchers as an effective solution to address the data scarcity issue. TL involves using a pre-trained CNN model for a new task. The model is initially trained on a specific dataset to learn features relevant to a particular task. Subsequently, it is adapted or fine-tuned for the new task, even if it belongs to a different domain. Another approach to overcoming data limitations is data augmentation, which increases the number of training images by applying transformations such as flipping, translation, zooming, and rotation. The key contributions of this paper include:

- Two deep lightweight CNN models, MobileNet and NASNetMobile, trained on the ImageNet dataset, are utilized for retinal disease classification from medical images.

- Transfer learning (TL) and data augmentation methods are utilized to overcome data limitations, improve feature representation, and reduce the likelihood of overfitting.

- The proposed models' performance is assessed using several metrics, such as accuracy, recall, precision, and F1-score, to evaluate their effectiveness in identifying retinal diseases.

## II. RELATED WORK

This section highlights recent studies that have employed artificial intelligence techniques in the diagnosis of retinal diseases.

[8] introduced a method to enhance the performance of optical coherence tomography (OCT) images using a pre-trained deep neural network (DNN). The authors proposed a strategy to resolve the problems posed by integrating networks built for natural images into the medical field. They altered the structure of GoogLeNet, ResNet, and DenseNet by deep custom-tailoring the convolutional structures to lessen the severity of the upper-level features images possess after the transfer learning.

[9] developed a deep CNN consisting of 18 convolutional layers and 3 fully connected layers to classify and stage diabetic retinopathy (DR) using fundus images. They employed a pre-processing stage involving image resizing and class-specific data augmentation. Using 5-fold and 10-fold cross-validation, the model achieved a validation accuracy of 88%-89%.

[10] proposed two frameworks for retinal image classification to detect maculopathy. The first framework used fuzzy preprocessing to enhance contrast followed by segmentation to extract blood vessels and the optic disc. Later histogram analysis was used to differentiate between normal and abnormal cases based on exudates. The second framework relied on (CNN) for automatic classification, the results showed the efficiency of the fuzzy preprocessing step.

[11] investigated the use of DL techniques for classifying retinal diseases using optical coherence tomography (OCT) images they employed five pre-trained DL, model-VGG-16, MobileNet, ResNet-50, Inception V3, and Xception- using a dataset consisting of 40,000 samples of OCT images.

[12] proposed a DL-based system to diagnose eye diseases using (CNNs).They employed three popular CNN architectures—VGG16, ResNet-50, and Inception-v3- and evaluated them on a large dataset of eye images from the internet. Inception-v3 performed the best with 97.08% accuracy.

[6] developed a lightweight CNN based on transfer learning to classify Diabetic Retinopathy (DR) and Diabetic Macular Edema (DME) severity. The proposed model addresses challenges related to deploying computationally intensive DL models on mobile or embedded devices with limited resources. By using ShuffleNetV2 as the base architecture, the model reduces the number of network parameters by approximately 28% and 5.5% compared to MobileNet V2 and ResNet50, respectively. The model's recognition speed improved from 73 to 40 milliseconds per image.

[13] conducted a study focused on the early diagnosis of diseases through the analysis of retinal blood vessels in fundus images. They proposed using deep learning (DL) -based classification with eight pre-trained CNN models along with enhancing transparency using Explainable AI tools like Grad-CAM, Score-CAM, and Layer CAM. The study also explored various architectures, including ResNet50V2, DenseNet121, Swin-Unet, and TransUNet.

[14] investigated the use of CNNs for classifying eye diseases and evaluated multiple pre-trained CNN architectures including VGG-16, VGG-19, ResNet-50, ResNet-152, and DenseNet-121, It was found that VGG-19 outperformed the other models achieving 95% accuracy.

[15] proposed a method for diagnosing eye disorders such as GLC and CAT using (CNNs) and (ANNs),The study showed that these networks achieved high accuracy in classifying conditions like GLC and DR.

## III. METHODS AND MATERIALS

### A. Description of the Dataset Used

The RFMiD (Retinal Fundus Multi-Disease Image Dataset) [16] was used to train and test the proposed models. It consists of 3,200 fundus images captured using three different types of fundus cameras, covering 46 disease categories, each representing a different retinal condition. Although the RFMiD dataset includes a large number of classes, this research focuses on two common diseases as well as healthy retinas. The two diseases are Diabetic Retinopathy (DR), which is caused by prolonged high blood sugar levels leading to damage in the small blood vessels of the retina, and Macular Hole (MH), a small break or defect in the macula, the central part of the retina responsible for sharp central vision. The dataset was split into two main sets: the training set and the test set. The data was randomly divided to ensure a balanced distribution of samples between the training and test sets. Table 1 shows the distribution of images in the dataset used in the work. Figure 1 shows examples from the dataset.

TABLE I. DATA DIVISION DETAILS OF NORMAL, DR, AND MH CLASSES

| Class name | No. of training images | No. of testing images |
|---|---|---|
| Normal | 401 | 134 |
| DR | 376 | 124 |
| MH | 312 | 104 |

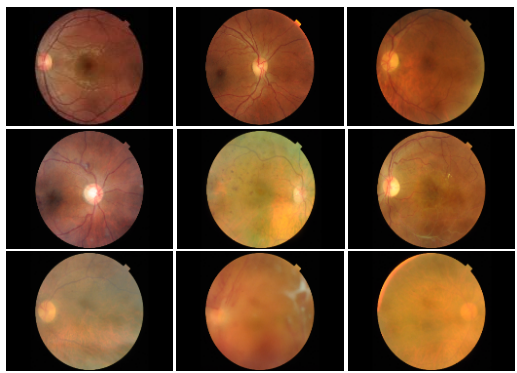

Fig.1. Examples from the RFMiD dataset. The first row shows healthy retinas, the second row shows DR, and the final row shows MH.

### B. Data Preprocessing

To ensure the compatibility of the datasets used in this work with the model's requirements, several essential preprocessing steps were carried out. These steps included:

- Image resizing, to reduce the dimension of the images into 224×224 to meet the requirement of the proposed models.

- Min-Max Normalization, to scale pixel values to the range [0,1] by dividing each pixel value by 255 to ensure consistency in data representation and improve model stability during training.

- Data Augmentation, to reduce the risk of overfitting and enhancing the model's performance when dealing

with unseen data by generating additional variations from the original images. The data augmentation process, including rotation by up to 30 degrees, horizontal and vertical shifts up to 20%, shearing transformations (up to 20%), zooming up to 20%, random horizontal flipping.

*C. CNN Architectures*

This work employed two lightweight and efficient CNN designed for mobile and embedded devices, MobileNet and NASNetMobile. MobileNet starts with fully convolutional layers with 32 filters and includes 19 residual bottleneck layers. The architecture consists of two modules, each with three layers, beginning and ending with a 1×1 convolutional layer. The second module acts as a fully connected layer with a depth of one. The ReLU activation is applied at various levels throughout the architecture. The key difference between the modules is their stride length: the first uses a stride of 1, while the second uses a stride of 2 [17].

NASNet Mobile begins with an initial step that includes a convolutional layer, which is then followed by the repeating Normal and Reduction Cells. Normal Cells keep the dimensions of the input while Reduction Cells resize with convolutional stride-2 downsampling. The model implements Depthwise Separable Convolutions to decrease the cost of computation without losing accuracy, and it ends with a Global Average Pooling layer and a Fully Connected layer with classification done through Softmax [18].

*D. The Proposed Model*

In this work, two lightweight Deep CNN networks, NASNetMobile and MobileNetV2, were used for the classification of retinal abnormalities. The two models are known to be highly efficient and perform well on the ImageNet dataset, thus making them suitable for use in medical imaging.

To adapt these models to the task of classifying retinal images into three categories: Normal, DR, and MH, modifications were made by removing the fully connected layers and replacing them with custom classification layers, as shown in Figure 2. These modifications aim to improve the models' ability to distinguish between different categories with higher accuracy. The added layers are:

- Global Average Pooling layer to convert the feature maps into a fixed-length feature vector by averaging each feature map.

- Flatten layer to convert the two-dimensional feature maps into a one-dimensional representation to facilitate classification.

- Dense layer - 1024 units with ReLU activation to capture intricate patterns within the feature vector, with ReLU activation enhancing the model's ability to learn nonlinear patterns.

- Dropout layer at 30% to reduce overfitting by deactivating some neurons during training, helping improve the model's generalization ability.

- Softmax layer to calculate the probabilities of belonging to each of the three categories, ensuring that the sum of the probabilities equals 1.

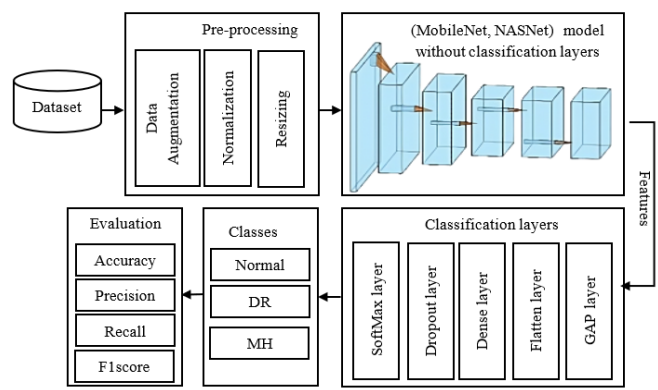

Fig.2. The block diagram of the proposed model.

IV. RESULTS

*A. Experment Setup and Training Options*

The proposed models were built using Python version 3.6, with the use of two open-source libraries, TensorFlow and Keras. The system used was a Dell Latitude 5420 laptop equipped with an Intel® Core™ i5-1135G7 processor (11th Gen) with a base frequency of 1.38 GHz and a maximum turbo frequency of 2.40 GHz, featuring 4 cores and 8 logical processors. It also includes 8GB of RAM and an Intel Iris Xe Graphics integrated GPU.

For training, several important hyperparameters were set to ensure optimal model performance. The learning rate was adjusted to 0.001 while the Adam optimizer was selected in order to balance the speed and accuracy of the updates done. The batch size was 32 for 30 epochs, and a dropout of 0.3 was included in an attempt to control overfitting. In addition, it was necessary to add a fully connected layer with 1024 neurons in the model to effectively learn the patterns in the data.

*B. Experment Results Using Proposed Models*

The (RFMiD) dataset was used as the training and testing data the proposed models for retinal disease classification. These models, MobileNetV2 and NASNetMobile, were trained and validated over several epochs. During this process, various performance metrics such as accuracy, recall, precision, and F1-score [19] were measured, the results of these models are summarized in Table 2.

MobileNetV2 scored the highest accuracy at 90.8%, followed by NASNetMobile with an accuracy of 89.5%. The recall metric, which measures the model's ability to correctly identify positive cases, was consistently high for both models with MobileNetV2 at 90.6% and NASNetMobile at 89.7%. In terms of precision, both models performed strongly, with MobileNetV2 achieving 90.6% and NASNetMobile slightly lower at 89.7%. The F1-score, which is a balanced measure of the model's accuracy, was 90.7% for MobileNetV2 and 89.1% for NASNetMobile. These results show that both models performed well in classifying retinal diseases, with MobileNetV2 slightly outperforming NASNetMobile overall.

TABLE II. THE EXPERIMENTAL RESULTS OF THE PROPOSED CNN MODELS ON RFMiD DATASET

| Model | Accuracy | Recall | Precision | F1-Score |
|---|---|---|---|---|
| MobileNetV2 | 90.8% | 90.6% | 90.6% | 90.7% |
| NASNetMobile | 89.5% | 89.7% | 89.7% | 89.1% |

Figure 3 shows the confusion matrix, the accuracy, and loss curves for both training and validation phases. The comparison between MobileNetV2 and NASNetMobile, shows that both models perform effectively in distinguishing between normal cases and diseases like DR and MH.

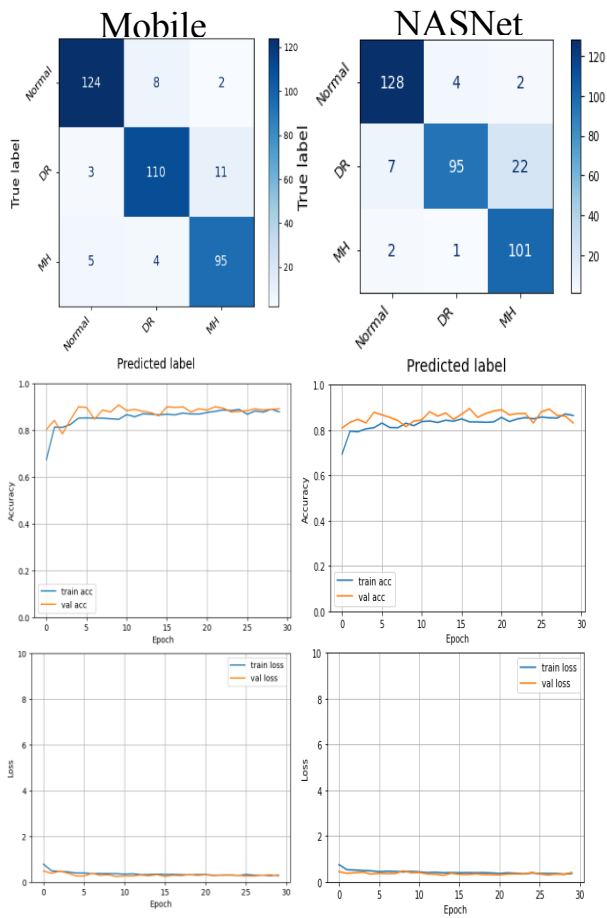

Fig.3. Confusion matrices, accuracy, and loss curves of the MobileNet and NASNetMobile models.

### C. Comparison of performance of proposed approach with existing methods

To evaluate the effectiveness of the proposed lightweight CNN models, MobileNetV2 and NASNetMobile, their performance was compared against state-of-the-art methods in the literature. The comparison, summarized in Table 3, is based on classification accuracy using the RFMiD dataset for retinal disease detection.

Our proposed approach utilizes lightweight CNN models, namely NASNet and MobileNetV2. NASNet achieved an accuracy of 89.5%, which is competitive with existing methods while maintaining a lower computational cost. MobileNetV2 outperformed all compared models, achieving an accuracy of 90.8%, making it the most effective approach among the examined methods. These results indicate that our lightweight CNN models particularly MobileNetV2 provide, an optimal balance between accuracy and computational efficiency.

TABLE III. COMPARISON OF THE PERFORMANCE OF THE PROPOSED MODELS WITH EXISTING METHODS.

| [Ref.], year | Method | Accuracy |
| --- | --- | --- |
| [20], 2020 | EfficientNetB3 | 90% |
| [2], 2022 | semi-supervised GANs | 87% |
| [21], 2024 | ViLReF | 84.82% |
| [22], 2024 | 20-layer CNN | 90.34% |
| [23], 2024 | Xception | 90.78% |
| Our Method | Lightweight CNN (NASNet) | 89.5% |
| Our Method | Lightweight CNN (MobileNetV2) | 90.8% |

## V. CONCLUSION

This study examined the application of Convolutional Neural Networks (CNNs) in detecting Diabetic Retinopathy (DR) and Macular Hole (MH), using fundus images. By employing MobileNetV2 and NASNetMobile, two efficient DL models, we achieved high classification accuracy while ensuring computational efficiency. MobileNetV2 outperformed NASNetMobile, achieving 90.8% accuracy, making it a promising solution for real-world medical applications. The results indicate that CNN-based models can enhance diagnostic accuracy, enabling early detection and timely intervention for retinal diseases. Future work could focus on expanding the dataset, integrating additional retinal conditions, and improving model designs for higher accuracy and efficiency.